\documentclass[mathleft]{an}
\usepackage[pdftex]{graphicx}
\usepackage{times}
\twocolumn
\begin{document}

\Pagespan{1}{8}
\Yearpublication{2010}%
\Yearsubmission{2010}%
\Month{}%
\Volume{}%
\Issue{}%
\DOI{}
\title{STK: A new CCD camera at the University Observatory Jena\thanks{Based on observations obtained with telescopes of the University Observatory Jena, which is operated by the Astrophysical Institute of the Friedrich-Schiller-University.}}

\author{M. Mugrauer\inst{1}\fnmsep
\thanks{Corresponding author: {markus@astro.uni-jena.de}}
\and  T. Berthold \inst{2}}
\titlerunning{STK: A new CCD camera at the University Observatory Jena}
\authorrunning{M. Mugrauer \& T. Berthold}
\institute{Astrophysikalisches Institut und Universit\"{a}ts-Sternwarte Jena, Schillerg\"{a}{\ss}chen 2-3, 07745 Jena, Germany \and Sternwarte Sonneberg\,/\,4pi Systeme GmbH, Sternwartestr. 32, 96515 Sonneberg, Germany}

\received{2010 Feb 10}
\accepted{2010 Feb 16}
\publonline{2010 Apr 19}

\keywords{instrumentation: detectors -- instrumentation: miscellaneous -- telescopes}

\abstract{The Schmidt-Teleskop-Kamera (STK) is a new CCD-imager, which is operated since begin of 2009 at the University Observatory Jena. This article describes the main characteristics of the new camera. The properties of the STK detector, the astrometry and image quality of the STK, as well as its detection limits at the 0.9\,m telescope of the University Observatory Jena are presented.}

\maketitle

\section{Introduction}

The University Observatory Jena is located close to the small village Gro{\ss}schwabhausen, west of the city of Jena. The Friedrich Schiller University operates there a 0.9\,m reflector telescope, which is installed at a fork mount (see, e.g., Pfau 1984)\nocite{pfau1984}. The telescope, as well as all its instruments are operated from a control room, which is located in the first floor of the observatory building directly beneath the telescope dome.

The 0.9\,m telescope can be used either as a Schmidt-camera, or as a Nasmyth telescope. In the Schmidt-mode ($f/D=3$) the telescope aperture is limited to \mbox{$D=0.6$\,m}, the aperture of the installed Schmidt-plate. In the Nasmyth-mode the full telescope aperture ${D=0.9}$\,m is used at  ${f/D=15}$. At the Nasmyth port of the telescope the spectrograph \mbox{FIASCO} (see Mugrauer \& Avila 2009)\nocite{mugrauer&avila2009} is installed, which also can be operated together with the CCD-camera CTK (see Mugrauer 2009)\nocite{mugrauer2009} for simultaneous spectro-photo\-metric monitoring of targets.

Already during the implementation phase of FIASCO at the 0.9\,m telescope the design study of a new CCD-imager for the Schmidt-focus of the 0.9\,m telescope was started. After the specification of its opto-mechanical design the, so called {S}chmidt-{T}eleskop-{K}amera (STK) was then built during 2008 by the company 4pi Systeme GmbH, under contract and after consulting the Astrophysical Institute and University Observatory Jena. First laboratory tests of the STK CCD-detector were then carried out at the end of 2008. Finally, at the begin of February 2009 the STK was installed in the tube of the 0.9\,m telescope and saw its first light at the University Observatory Jena. Afterwards, the instrument was dismounted again as some modifications of its filter system (in particular software updates and laboratory tests of the filter system hardware) were necessary. End of June 2009 the STK was then reinstalled in the Schmidt-focus of the 0.9\,m telescope for its first science operation run, which last until mid of November 2009.

During the first months of operation at the 0.9\,m telescope the properties of the STK detector were determined (dark current, bias level, linearity), and several photo- and astrometric observations were carried out with the new camera to determine its image quality, astrometry, as well as its detection limits. Beside the instrument characterization, the STK was already used in several scientific projects, mainly photometric monitoring programs to study the variability of stars in young stellar clusters, follow-up observations of long-periodical variable stars, as well as of stars with transiting planets.

In this paper we present the main characteristics of the STK. In the second section the opto-mechanical design, as well as all individual components of the new camera are described in detail. Section 3 summarizes all properties of the
STK detector. The astrometry and image quality of the new CCD-imager are discussed then in Sect. 4. The STK detection limits at the 0.9\,m telescope of the University Observatory Jena are presented in Sect. 5. The illumination effect induced by the STK shutter in short integrated images is shown in Sect. 6. Finally, the first light observations of the STK at the 0.9\,m telescope of the University Observatory Jena are presented in the last section.

\section{STK components}

\begin{figure}[t]
\centering
\resizebox{0.9\hsize}{!}{\includegraphics[]{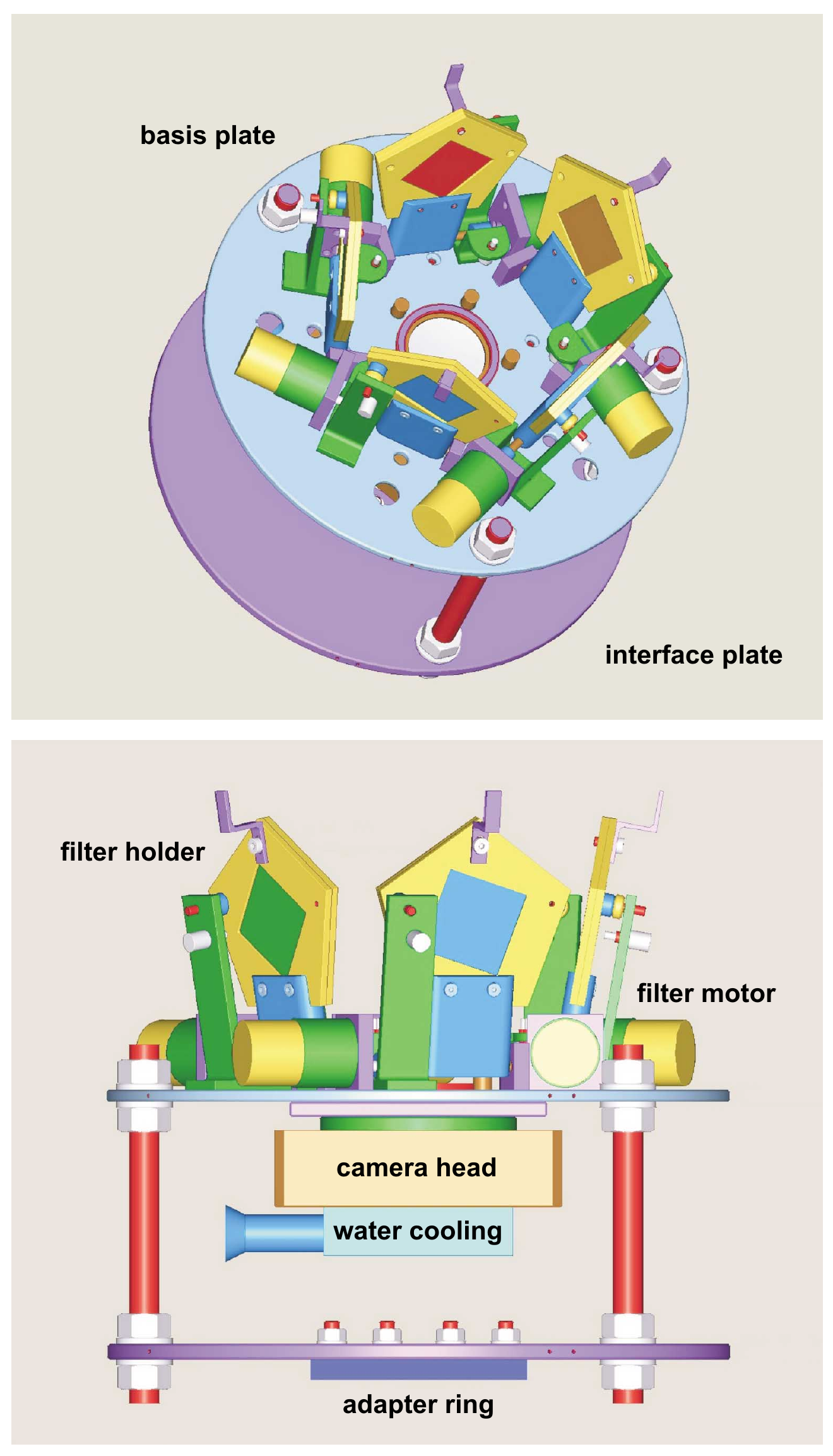}} \caption{The \emph{top panel} shows the front-view of the STK opto-mechanical design, without its dust protection cover. The basis plate, on which the filter system of the STK and its field-flattener optics are installed, is located in the front with the interface plate behind. The \emph{bottom panel} shows the side-view of the STK opto-mechanical design. The individual filter holders and motors of the STK filter system, the STK camera head with its water cooling unit on the basis plate, as well as the adapter ring on the bottom side of the interface plate are indicated.} \label{layout}
\end{figure}

The opto-mechanical design of the STK is shown in Fig.\ref{layout}. The camera is made up of two circular plates, the so-called basis and interface plate, each with an outer diameter of 334\,mm. Both plates are mounted parallel to each other with a separation of 135\,mm, and are orientated exactly perpendicular to the optical axis of the 0.9\,m telescope. The basis plate carries on its top side the filter system of the STK, as well as its field-flattener optics. The STK camera head with its water cooling unit are installed on the bottom side of this plate. With the adapter ring, which is mounted on the bottom side of the STK interface plate, the camera is installed at the focusing unit of the 0.9\,m telescope. The electronically controllable focusing unit is integrated in the secondary mirror support structure of the 0.9\,m telescope and is operated from the control room of the observatory.

Because the STK is installed in the tube of the 0.9\,m telescope a specific design for the STK filter system is  needed, in order to minimize its outer diameter, which limits the obstruction of the free aperture of the 0.9\,m telescope. Due to the special design of the STK filter system the outer diameter of the instrument measures only 334\,mm, which effectively reduces the diameter of the free telescope's aperture in the Schmidt-mode ($D=0.6$\,m) only to $D=0.5$\,m. The STK filter system is equipped with five filter holders in which quadratic filters each with an edge length of 50\,mm and a thickness of 5\,mm can be installed. The filter holders are driven by their accessory motors, which move the individual filter holders into the light beam of the 0.9\,m telescope. Beside the observation with filters, the STK filter systems allows also imaging without filters for maximal sensitivity. The STK filter system is permanently equipped with four filters, namely the $B$-, $V$-, $R$-, and $I$-band filters from the Bessel standard system (\cite{bessel1990}). The fifth filter holder is used as a blank for dust protection of the instrument, but can also be equipped with additional filters, e.g. the $z'$-band filter from the SDSS system (\cite{fukugita1996}), which is presently available.

In order to correct the curvature of the focal plane of the 0.9\,m telescope in its Schmidt-focus the STK is equipped with a field-flattener optics, which is installed directly in front of the STK detector. The STK field-flattener is a
aspheric-concave lens, which is built out of the low dispersive Schott glass N-FK5, to suppress its chromatic abberation. With a diameter of 58\,mm the lens is sufficiently large to avoid any kind of vignetting of the telescope's light beam. The STK image quality achieved with its field-flattener optics in the Schmidt-focus of the 0.9\,m telescope is discussed in detail in Sect. 4.

\begin{figure}[h]
\centering
\resizebox{0.9\hsize}{!}{\includegraphics[]{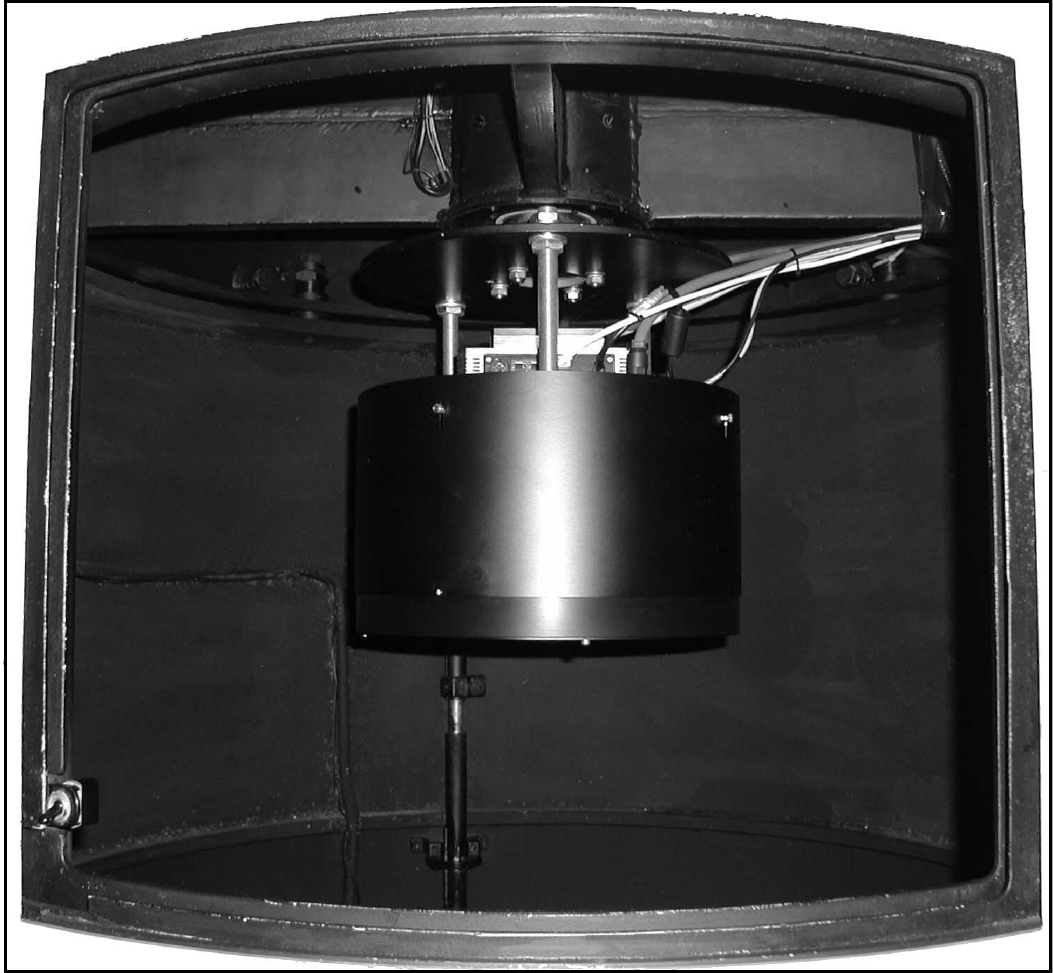}} \caption{This image shows the STK installed in the Schmidt-focus of the 0.9\,m telescope, ready for its night time operation. The filter-system of the camera, as well as its field-flattener optics are both covered by the black cylindric dust protection cover. The STK is mounted on the focusing unit of the telescope with the adapter ring, which is installed on the bottom side of the interface plate, as it is illustrated in Fig.\,\ref{layout}. } \label{stkpicture}
\end{figure}

All optical components of the STK are protected against dust pollution by a cylindric protection cover, which is  mounted on the top side of the basis plate and encircles the STK filter system and the field-flattener optics of the
instrument. The protection cover exhibits a height of 200\,mm and possesses a circular cutout on its front-side with a diameter of 130\,mm, large enough to avoid any kind of vignetting of the light beam of the 0.9\,m telescope. Figure\,\ref{stkpicture} shows the STK with its installed dust protection cover in the Schmidt-focus of the telescope.

The STK camera head contains the CCD-detector, its temperature control and stabilization unit, as well as the detector readout electronics. The STK detector is installed in an insulated cell with an entrance window in front of the detector. The detector cell is dried and filled with argon gas for suppression of humidity. An iris shutter with five shutter blades is installed in front of the entrance window of the detector cell. The shutter can be opened for integration times down to 0.03\,s.

Due to its operation in the Schmidt-focus of the 0.9\,m telescope, i.e. inside of the tube of the telescope, classical air cooling of the STK is not possible, as it would rapidly increase the air temperature inside the tube of the telescope, which would then significantly degenerate the image quality of the instrument, because of tube seeing. In order to avoid the warming-up of the air in the tube of the telescope and therefore to guarantee a good image quality during STK operation, the instrument is operated with a water cooling system. A water cooling aggregate, which is installed in the dome of the observatory just a few meters away from the telescope, is connected by two water pipes with the backside of the STK camera head. The water cooling aggregate permanently pumps water through the water cooling unit of the STK camera head, where it effectively removes energy from the heat sink of the STK detector. The water cooling aggregate is temperature stabilized (0.1\,K accuracy) and water temperatures between 14 and 22$^{\circ}$C can be chosen. In order to avoid condensation on the STK housing, on all water pipes, as well as in the water cooling aggregate itself, the temperature of the cooling water always has to be chosen at least 2\,K above the current dew-point temperature.

\section{STK detector}

\begin{figure}[t]
\centering
\resizebox{0.9\hsize}{!}{\includegraphics[]{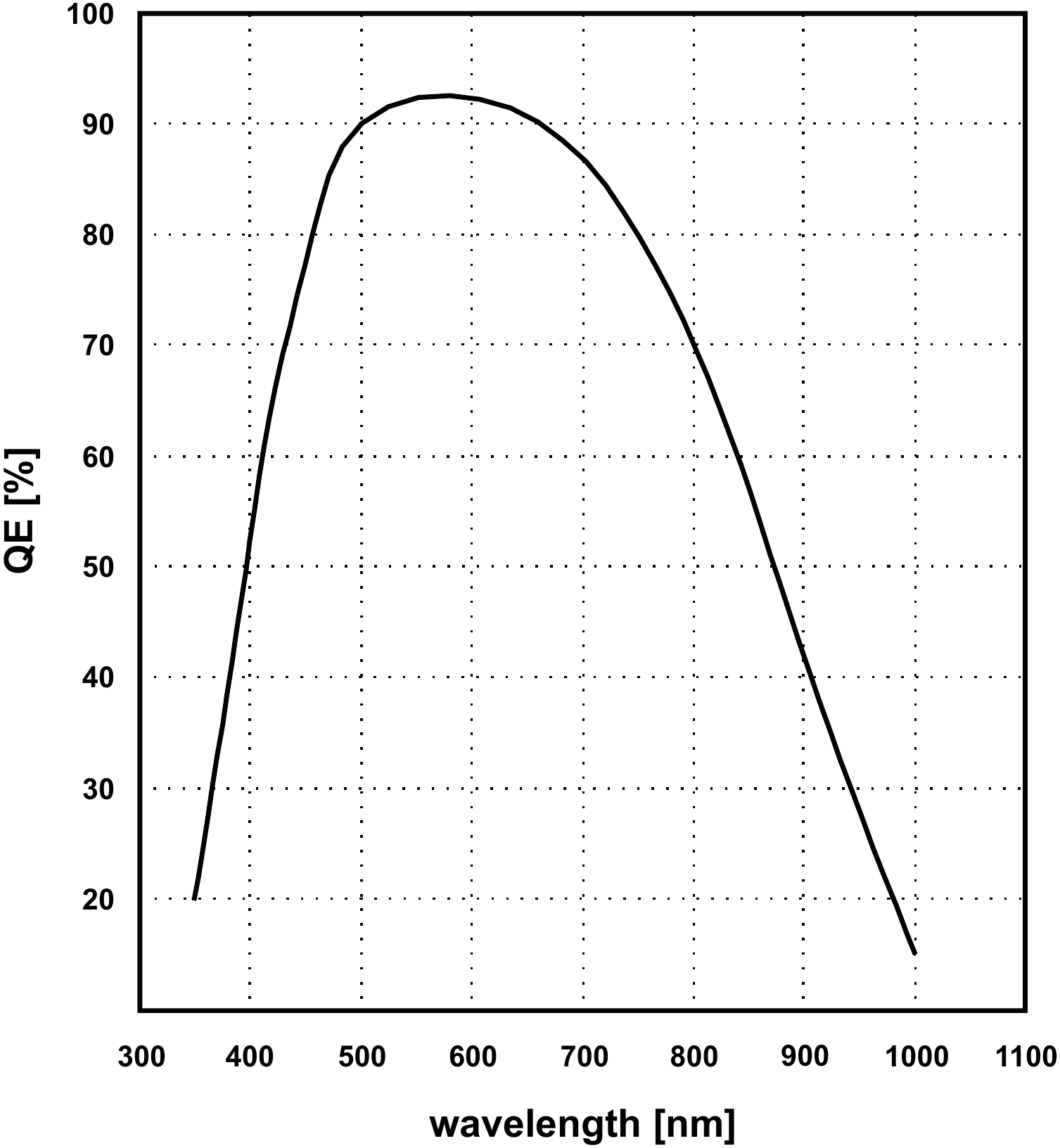}} \caption{The quantum efficiency of the STK detector for a range of wavelength, measured at a detector temperature of $-$20\,$^{\circ}$C. Because of its back-illumination and special coating the STK detector exhibits a high quantum efficiency, which remains above 70\,\% between 430\,nm and 800\,nm, and reaches its maximum of 92\,\% at 580\,nm.} \label{qe}
\end{figure}

The STK detector is a midband coated E2V CCD-42-40 of grade 1, which is sensitive in the wavelength range between 200 and 1060\,nm. Due to its back-illumination and special coating the STK detector exhibits a high quantum efficiency, which remains above 70\,\% between 430\,nm and 800\,nm, and reaches its maximum of 92\% at a wavelength of 580\,nm. The quantum efficiency of the STK detector is shown in Fig.\,\ref{qe} for a range of wavelength.

The STK detector measures 27.65\,mm\,$\times$27.65\,mm and is composed out of 2048$\times$2048 pixels, each with a size of 13.5\,$\mu$m\,$\times$13.5\,$\mu$m. In order to monitor the bias level of the STK detector an overscan region of 50
detector columns is always read out, resulting in a total image size of 2098$\times$2048 pixels. The STK detector is operated in the advanced inverted mode, which significantly reduces the dark current of the detector, and therefore allows its operation on a low-noise level even at higher detector cooling temperatures above $-50\,^{\circ}$C.

The STK detector is Peltier cooled and its temperature is regulated and stabilized by the detector cooling electronics, and a temperature difference up to 45\,K below the chosen water cooling temperature is reachable. In order to minimize the thermal stress of the STK detector during its cool-down or warm-up phase, the detector cooling electronics changes only slowly the temperature of the detector over time. Hence, for the cool-down or warm-up of the STK detector about 30\,min are needed until the chosen detector temperature is reached and stabilized by the detector cooling electronics.

\begin{figure}[t]
\centering
\resizebox{0.9\hsize}{!}{\includegraphics[]{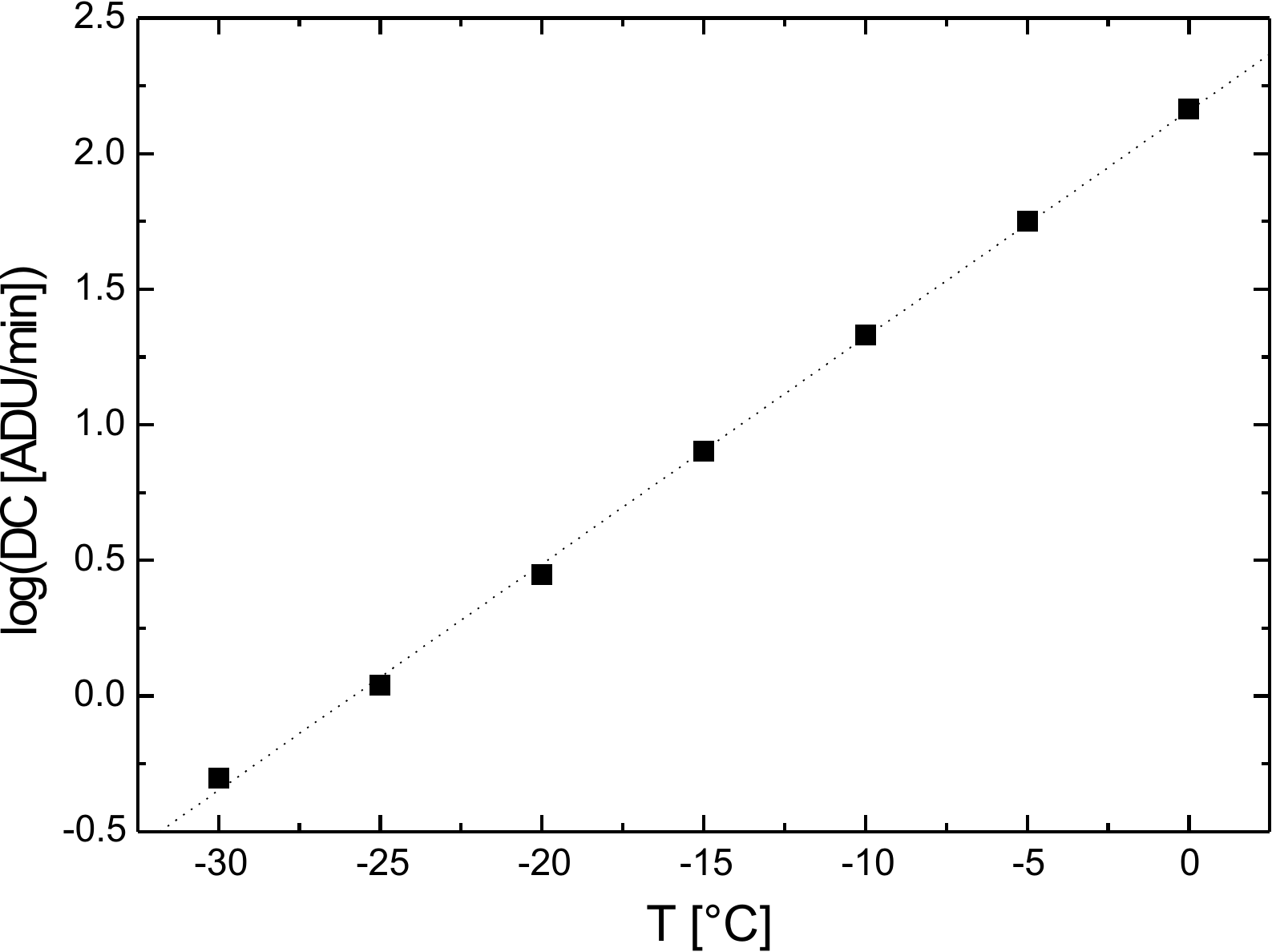}} \caption{The dark current of the STK detector for a range of detector cooling temperature. The dark current scales exponentially with the cooling temperature of the STK detector and remains low up to the critical cooling temperature of $-25\,^{\circ}$\,C. Therefore, the STK should always be operated at detector cooling temperatures below this critical temperature limit.} \label{darktemp}
\end{figure}

The dark current and the bias level of the STK detector, as measured in dark and bias series for different detector cooling temperatures from $-30$ up to $0\,^{\circ}$\,C, are summarized in Table\,\ref{table_darkbiastemp}. The dependency of the dark current on the chosen detector cooling temperature is also illustrated in Fig.\,\ref{darktemp} in the logarithmic scale. As expected for a CCD-sensor, the measured dark current scales exponentially with the detector cooling temperature. The logarithmic dependency of the dark current (DC in the unit of ADU/min) on the detector cooling temperature ($T$ in the unit of $^{\circ}$\,C) can be described well with the linear fit
\begin{flushleft}
$\log (\mathrm{DC})= 2.1584\pm0.0212 + (0.08355\pm0.00118) \times T$
\end{flushleft}
with a correlation coefficient $R=0.99951$.

\begin{table}[t!] 
\caption{The dark current for 1\,min of integration time, as well as the bias level of the STK detector for a range of detector cooling temperature. The dark current remains on a low level for cooling temperatures below about $-25\,^{\circ}$C, the so-called critical cooling temperature. Operation of the STK above this temperature limit should be avoided, due to a significantly increased dark current.} \label{table_darkbiastemp}
\begin{tabular}{ccc}\hline\noalign{\smallskip}
Temperature & Dark Current & Bias \\[1.5pt]
$[^{\circ}$C] & [ADU/min] & [ADU]\\[1.5pt]
\hline\noalign{\smallskip}
$-30$   & $\,\,\,\,\,\,0.5 \pm 0.1$     & $1223 \pm 1$\\
$-25$   & $\,\,\,\,\,\,1.1 \pm 0.2$     & $1224 \pm 1$\\
$-20$   & $\,\,\,\,\,\,2.8 \pm 0.2$     & $1229 \pm 1$\\
$-15$   & $\,\,\,\,\,\,8.0 \pm 0.3$     & $1238 \pm 1$\\
$-10$   & $\,\,\,21.5 \pm 0.4$          & $1261 \pm 1$\\
$\,\,\,-5$   & $\,\,\,56.2 \pm 1.0$     & $1322 \pm 1$\\
$\,\,\,\,\,\,\,\,0$ & $145.8 \pm 1.4$   & $1471 \pm 3$\\[1.5pt]
\hline
\end{tabular}
\end{table}

The detector cooling electronics stabilizes the temperature of the STK detector on the 0.1\,K level. Hence, the variation of the dark current of the STK detector is negligible during night time operation. While the dark current of the STK detector remains low below a critical cooling temperature of $-25\,^{\circ}$\,C, it significantly increases at higher detector temperatures. Therefore, it is recommended to operate the camera always below its critical temperature limit of $-25\,^{\circ}$\,C. The STK water cooling system guarantees operation below this critical temperature limit during all nights of a year. In cold winter nights detector cooling temperatures down to $-31\,^{\circ}$C can be chosen, while in typical warm summer nights the STK can be operated at higher temperatures between $-30$ and $-25\,^{\circ}$C, which also yield low dark currents.

Beside the dark current also the detector bias level  shows a dependency on the chosen detector temperature  (see Table\,\ref{table_darkbiastemp}). However, this effect is just significant for detector cooling temperatures above the critical temperature limit, a range of temperature in which the instrument generally should not be operated.

\begin{figure}[t]
\centering
\resizebox{0.9\hsize}{!}{\includegraphics[]{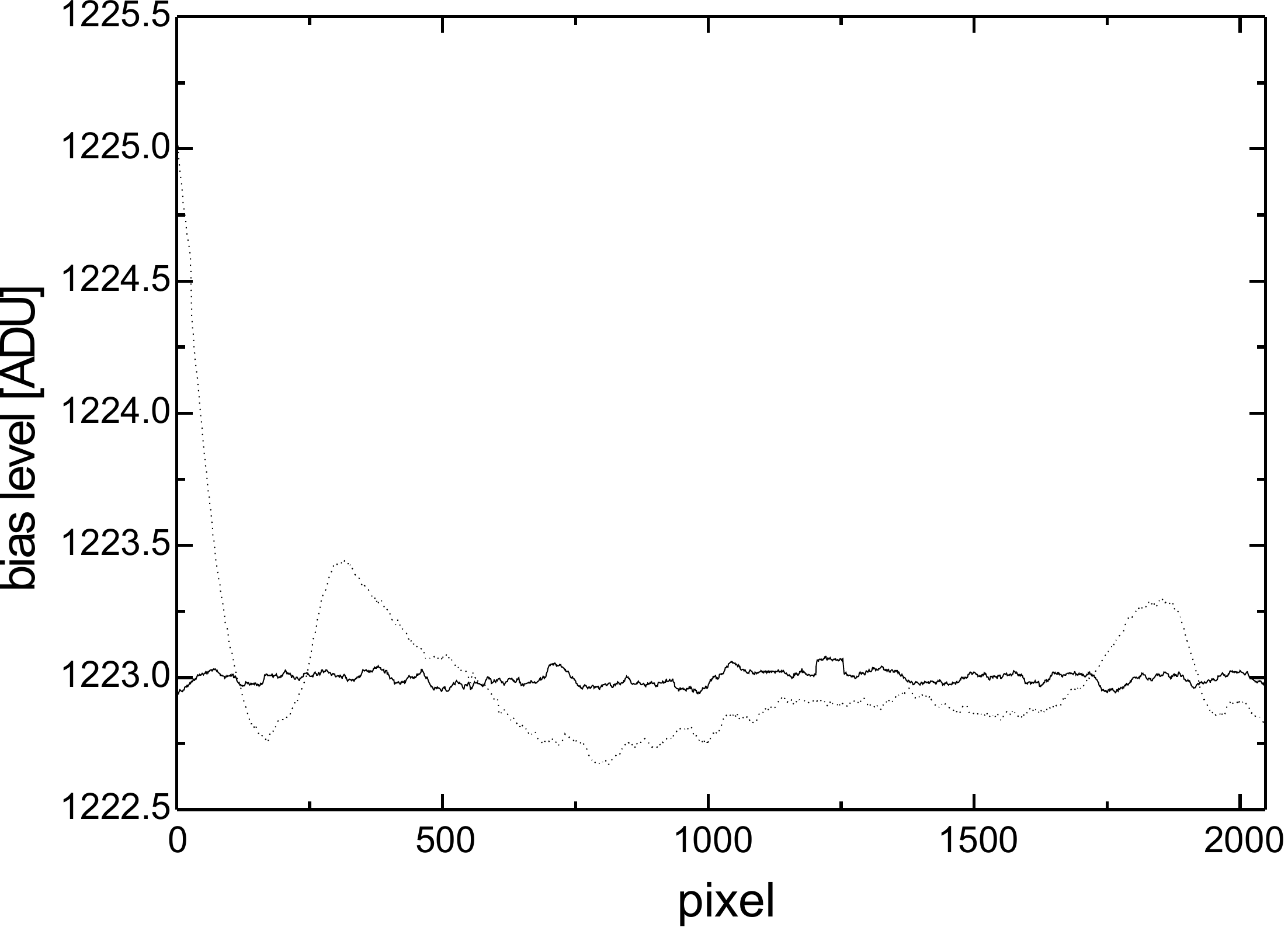}} \caption{The bias level measured over the whole STK detector at a detector temperature of $-30\,^{\circ}$. The solid line shows the bias level along the vertical direction for the average of all detector columns, while the dotted line illustrates the bias level along the horizontal direction for the average of all detector rows. The bias level of the STK detector is found to be highly stable on the 1\,ADU level over its whole area. Only at the left border of the STK detector \mbox{($x<75$\,pixels)} the bias level increases up to 2\,ADU relative to the average bias level, induced by the detector readout electronics.} \label{bias}
\end{figure}

Figure\,\ref{bias} shows the variation of the bias level over the  whole area of the STK detector. The bias level is found to be highly stable over the whole detector with a deviation of only 1\,ADU. While there is no significant variation of the bias level along detector columns, small changes of the bias level in the order of less than 1\,ADU are found along the detector rows. Only for detector columns at the left border of the STK detector ($x<75$\,pixels) an increase of the bias level up to 2\,ADU relative to its average is measured, induced by the detector readout electronics. Hence, any kind of strong trends of the bias level over the whole STK detector can clearly be ruled out. Furthermore, no significant changes of the bias level are detected during several consecutive observing nights, at constant detector and water cooling temperatures.

Although the STK detector exhibits a high stability of its dark current and bias level it is recommended to take dark frames once in each observing night for all used integration times to achieve calibration images of optimal quality for obtained science data.

\begin{figure}[t]
\centering
\resizebox{0.9\hsize}{!}{\includegraphics[]{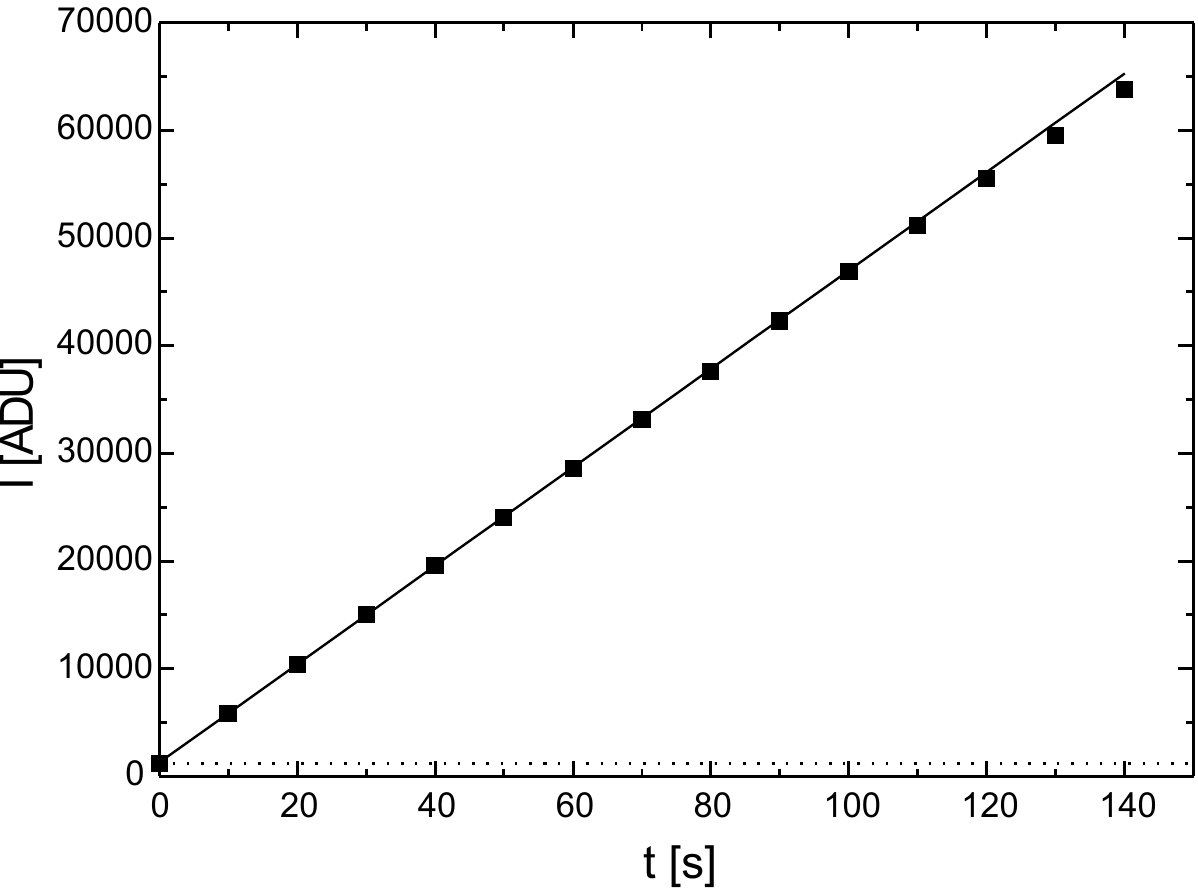}} \caption{The linearity of the STK detector, as measured in several domeflat images. The detector intensities $I$ of domeflats, taken at constant flat illumination, are plotted for a range of integration time. A quite linear dependency (straight line) between the integration time and the detector intensity $I$ is found for the whole dynamic range of the STK detector from its bias level at about 1220\,ADU (dotted line) up to the digitization limit of the STK readout electronics.} \label{linearity}
\end{figure}

In order to measure the linearity of the STK detector domeflats with different integrations times were taken at a constant flat illumination (see Fig.\,\ref{linearity}).

\begin{figure}[t]
\centering
\resizebox{0.9\hsize}{!}{\includegraphics[]{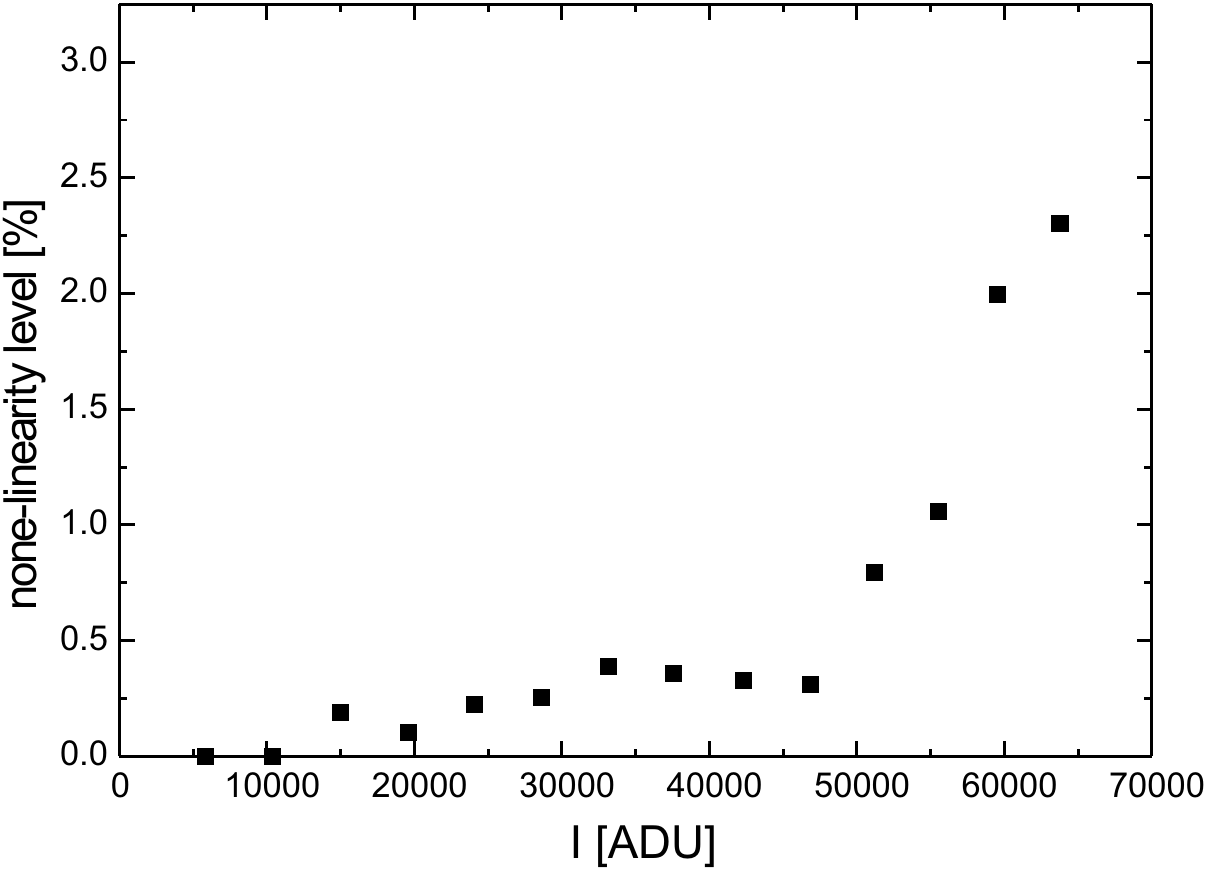}} \caption{This figure shows the none-linearity level of the STK detector, measured for a range of detector intensities. The STK detector is highly linear over its full dynamic range from the bias level at about 1220\,ADU to the upper limit of the ADC of the STK readout electronics. In particular, for detector intensities below 50\,000\,ADU the detector none-linearity level is smaller than 1\,\%, and only slightly increases up to 2.3\,\% at 64\,000\,ADU, i.e. close to the limit of the ADC of the STK readout electronics.} \label{linearitylevel}
\end{figure}

The STK detector is highly linear over its full dynamic range, from the bias level at about 1220\,ADU to the digitization limit of the readout electronics (65\,535\,ADU). The none-linearity level, i.e. the deviation of the slope compared to the slope measured for detector intensities up to 10000\,ADU is shown in Fig.\,\ref{linearitylevel}.

The STK detector exhibits a high level of linearity below an intensity of 50\,000\,ADU (none-linearity level smaller than 1\,\%). For higher detector intensities the none-linearity level increases slightly and reaches about 2.3\,\% at  64\,000\,ADU, i.e. close to the digitization limit of the STK readout electronics. Hence, precise photometric measurements are feasible with the STK, spanning the complete dynamic range of the instrument.

The analog-digital-converter (ADC) of the STK readout electronics works at 16\,bit, i.e. the file size of an individual STK image (2098$\times$2048 pixels, overscan included) is about 8.2\,MB. The STK detector is read out with a frequency of 700\,kpixel/s, followed by the data transfer to the camera control computer, which is done at USB2 standard. The measured total time for the full detector readout and data transfer of one STK image is about 11.5\,s.

Due to the operation of the STK detector in the advanced inverted mode its full well capacity is slightly reduced to about 100\,000\,e$^{-}$. Thereby, the gain (1.2\,${\rm e}^{-}/\rm ADU$) of the readout electronics is adjusted in a way that the ADC of the detector readout electronics digitized only the highly linear range of the STK detector. The typical read noise induced by the readout electronics was measured to be 8\,${\rm e}^{-}$.

\section{STK astrometry and image quality}

\begin{figure}[t]
\centering
\resizebox{0.9\hsize}{!}{\includegraphics[]{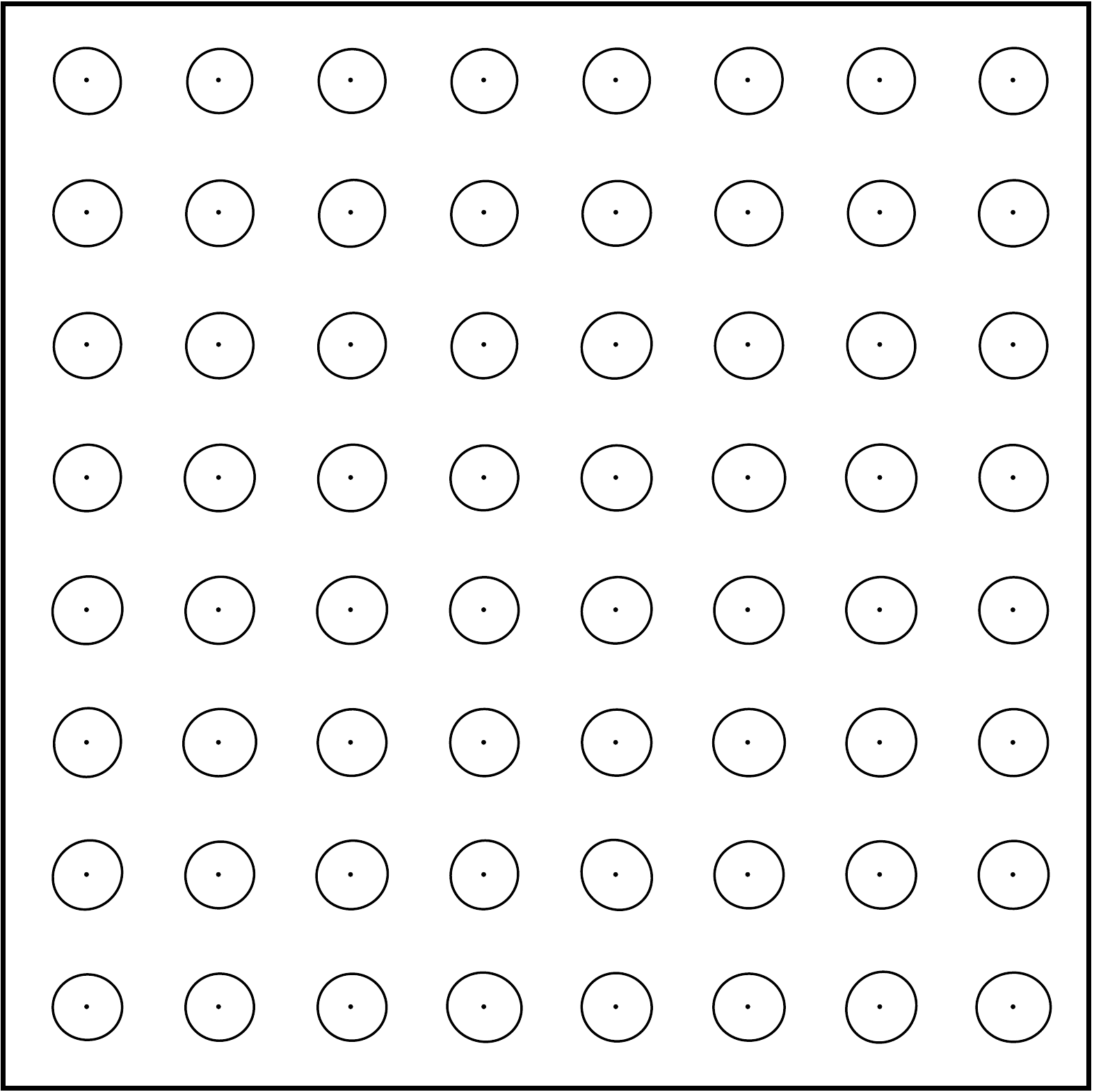}} \caption{This image shows the FWHM and shape of the STK-PSF, dependent on the position on the STK detector, as measured through the STK $V$-band filter. The smallest FWHM of the STK-PSF is reached in the center of the STK detector where it measures 2\,pixels. The PSF of the camera is well radial symmetric over the whole STK field of view and does not significantly change from the detector center to its outer border.}
\label{iqual1}
\end{figure}

With its 13.5\,$\mu$m pixels the STK detector exhibits a pixel scale of 1.546\,arcsec/pixel in the Schmidt-focus of the 0.9\,m telescope, which yields the STK field of view of  52.8$'\times$52.8$'$. The pixel scale of the STK detector was measured in several images taken at different telescope orientations. In order to derive the pixel scale, as well as the position angle of the STK detector, all images were astrometrically calibrated with reference sources from the 2MASS point source catalogue (\cite{skrutskie2006}). Furthermore, with the measured pixel scale of the STK detector and its known pixel size the focal-length of the Schmidt-focus  could be determined to be \mbox{$f=1801.3\pm0.3$\,mm}. The STK is mounted at a fixed orientation in the tube of the 0.9\,m telescope with a detector position angle of about
$-1.5\,^{\circ}$, i.e. the camera is rotated by about $1.5\,^{\circ}$ from north to west. The astrometry of the STK is summarized in Table\,\ref{table_astro} for both observing epochs in which the instrument was used, so far.

\begin{table}[t!]
\caption{This table summarizes the STK astrometry for both observing epochs in which the instrument was used, so far. The pixel scale PS and the position angle PA of the STK detector are listed.} \label{table_astro}
\begin{tabular}{ccc}\hline\noalign{\smallskip}
Epoch & PS & PA\\[1.5pt]
[mm/yyyy] & $[$arcsec/pixel$]$ & $[^{\circ}]$\\[1.5pt]
\hline\noalign{\smallskip}
02/2009 & $1.5459\pm0.0003$ & $-1.50\pm0.05$\\
07/2009 & $1.5463\pm0.0004$ & $-1.60\pm0.04$\\[1.5pt]
\hline
\end{tabular}
\end{table}

Although the STK was dismounted from the telescope between the first light run in February 2009 and its first science operation run, which started at the begin of July 2009, the astrometry of the instrument remains highly stable and no significant changes are detected between both observing epochs.

In order to determine the image quality over the full STK field of view, several images were taken, which contain hundreds of stars, distributed over the whole STK detector. For all detected point like sources in all images, the full-width-half-maximum (FWHM), as well as the ellipticity of the STK point-spread-function (PSF) was measured. Thereby, the STK field of view was split in cells, which measure each 256$\times$256 pixels. The average of the FWHM and the ellipticity of all point like sources, which are imaged in each cell, were derived. This yields the averaged FWHM and ellipticity of the STK-PSF for each cell, and hence the image quality over the whole STK field of view. The image quality of the STK, as determined in the V-band after optimized focusing (FWHM of 2\,pixels in the center of the STK detector), is shown in Fig.\,\ref{iqual1}.

\begin{figure}[t!]
\centering
\resizebox{0.9\hsize}{!}{\includegraphics[]{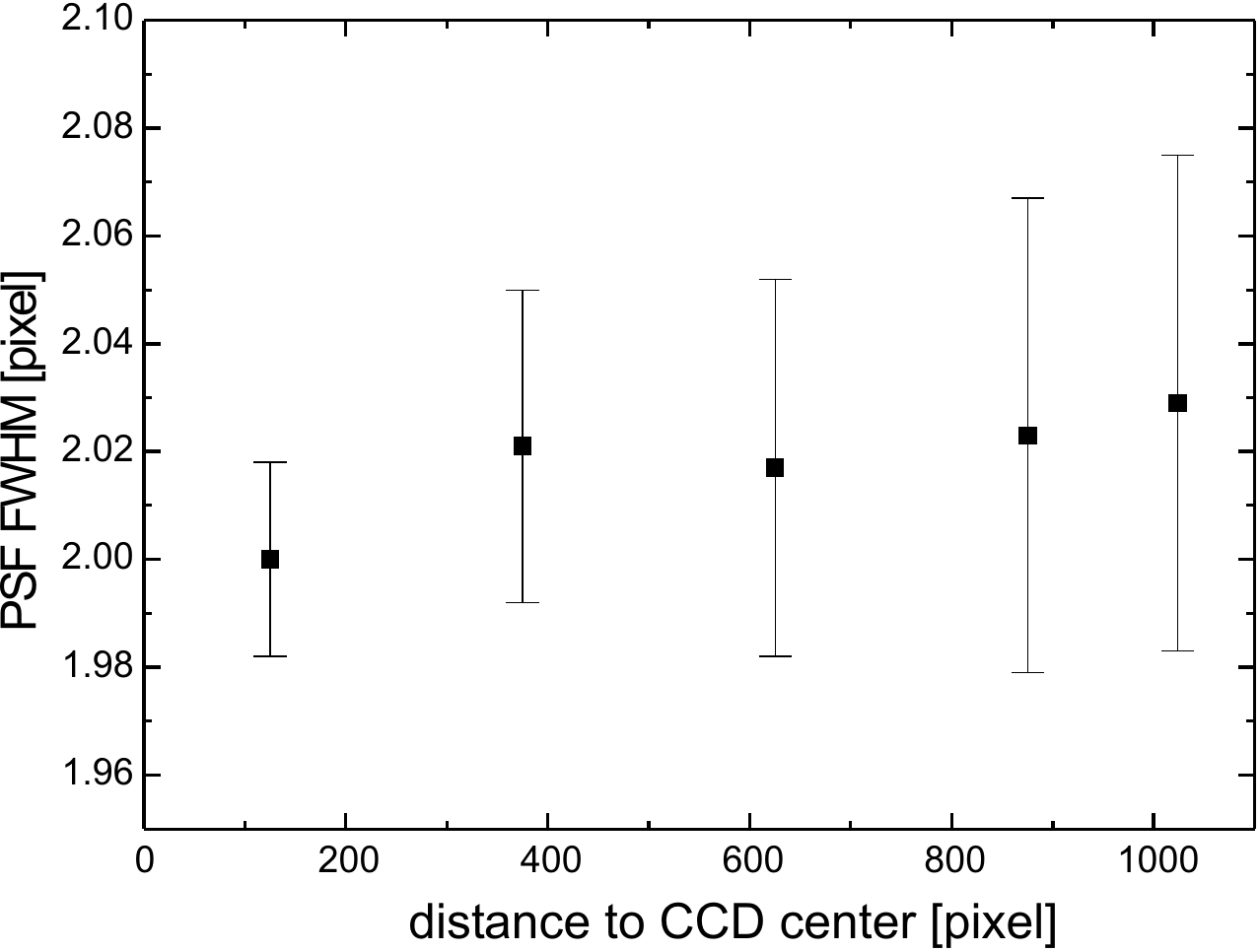}} \caption{The FWHM of the STK-PSF for a range of distance relative to the center of the STK detector. In order to determine the radial variation of the PSF-FWHM, all detector cells, as shown in Fig.\,\ref{iqual1}, are grouped dependant on their distance to the center of the STK detector. The average, as well as the scatter of the PSF-FWHM is derived for distance-bins of 250\,pixels. The FWHM of the STK-PSF does not show any significant variation over the whole STK field of view and is stable on the 2\,\% level. The scatter of the PSF-FWHM only slightly increases from 1 to 2\,\% from the center to the border of the STK detector.} \label{iqual2}
\end{figure}

\begin{figure}[t!]
\centering
\resizebox{0.9\hsize}{!}{\includegraphics[]{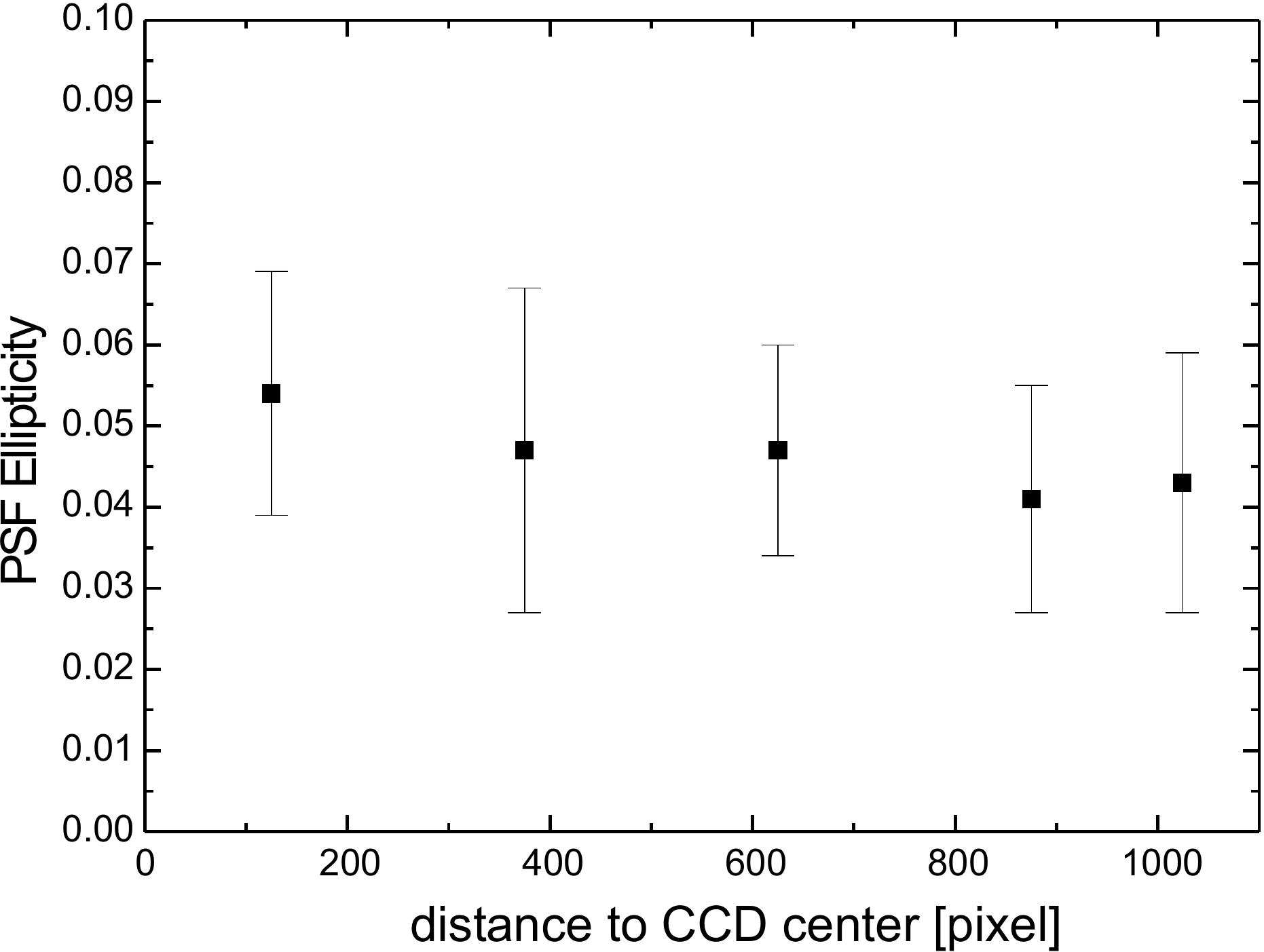}} \caption{The ellipticity of the STK-PSF for a range of distance relative to the center of the STK detector. In order to determine the radial variation of the PSF ellipticity, all detector cells, as shown in Fig.\,\ref{iqual1}, are grouped dependant on their distance to the center of the STK detector. The average, as well as the scatter of the PSF-ellipticity is derived for distance-bins of 250\,pixels. The ellipticity of the STK-PSF is determined to be $5\pm2$\,\% over the whole STK detector and does not show any significant variation within the field of view of the instrument.} \label{iqual3}
\end{figure}

The averaged FWHM and ellipticity of the STK-PSF, dependant on the distance to the center of the STK detector (pixel [1024,\,1024]), are plotted in Figs.\,\ref{iqual2} and \ref{iqual3}. Thereby, cells are grouped in bins of 250\,pixels dependant an their separation to the center of the STK detector. Within a distance smaller than 250\,pixels around the center of the STK detector the averaged PSF-FWHM is 2\,pixels with a scatter of only 1\,\%. The PSF-FWHM only slightly increases up to 2.03\,pixels with a scatter of 2\,\% at the outer edge of the STK detector. The overall ellipticity of the STK-PSF is found to be $5\pm2$\,\%. The STK image quality was also determined for all other permanently installed filters of the instrument ($B$, $R$, $I$) and is consistent with the measurements shown here for the $V$-band. Hence, the instrument exhibits a very good image quality with a highly stable PSF of radial symmetric shape over its whole field of view.

Due to the slightly different thickness of the individual filters, as well as the chromatic abberation of the STK field-flattener optics, the focal plane of the instrument slightly varies dependent on the chosen filter. In order to check for focus stability the FWHM of the STK-PSF was measured in several STK images during focus series. Thereby, the camera focus was optimized in the $V$-band with an averaged PSF-FWHM of 2\,pixels in the central part of the STK field of view. For the same focus position the FWHM of the STK-PSF was then measured with all other permanently installed STK filters. An increase of the PSF-FWHM is found to be in average 2.24\,pixels in $B$-, 2.07\,pixels in $R$-, and 2.46\,pixels in the $I$-band with a typical scatter of 1\,\%. Hence, the FWHM of the STK-PSF varies up to about 20\,\% dependant on the used filter. Therefore, it is recommended to focus the camera after a filter change in order to achieve always an optimal image quality. In particular, when changing from filter-free imaging to imaging with filters a strong de-focusing of the instrument (due to the thickness of the used filters) occurs, which has to be compensated. Precise focusing of the STK is easily and quickly doable from the control room of the observatory, using the focusing unit of the 0.9\,m telescope.

\section{STK detection limits}

\begin{table}[h!]
\caption{The STK detection limits (${\rm S/N}=3$) for 1\,min of integration time at intermediate elevations of about 45\,$^{\circ}$ (airmass 1.4) during dark time (no moon). The limits were determined after an optimized focusing of the camera in each filter.}
\label{stklimits}\begin{tabular}{cccc}\hline\noalign{\smallskip}
$B$ & $V$ & $R$ & $I$\\
$[$mag$]$ & $[$mag$]$ & $[$mag$]$ & $[$mag$]$\\[1.5pt]
\hline\noalign{\smallskip}
19.1$\pm$0.1 & 19.2$\pm$0.1 & 19.2$\pm$0.2 & 18.6$\pm$0.1 \\[1.5pt]
\hline
\end{tabular}
\end{table}

The detection limits of the STK at the 0.9\,m telescope were determined with several standard stars, which were  observed at intermediate elevations of about 45$^{\circ}$ (airmass 1.4) during dark time (no moon). The
integration time for each standard was always 1\,min with an optimized instrument focusing for each filter (FWHM of 2\,pixels). The obtained detection limits for the permanently installed STK filters ($B$, $V$, $R$, $I$) from the Bessel standard system (Bessell 1990) are summarized in Table\,\ref{stklimits}. After 1\,min of integration time stars fainter than the 19{th} magnitude are detectable with the STK at ${\rm S/N}=3$ in the $B$-, $V$-, and $R$-band. In the $I$-band stars down to the 18.6{th} magnitude can be imaged with the STK at ${\rm S/N}=3$ after 1\,min of integration time.

\section{Shutter illumination effect}

\begin{figure}[t!]
\centering
\resizebox{0.9\hsize}{!}{\includegraphics[]{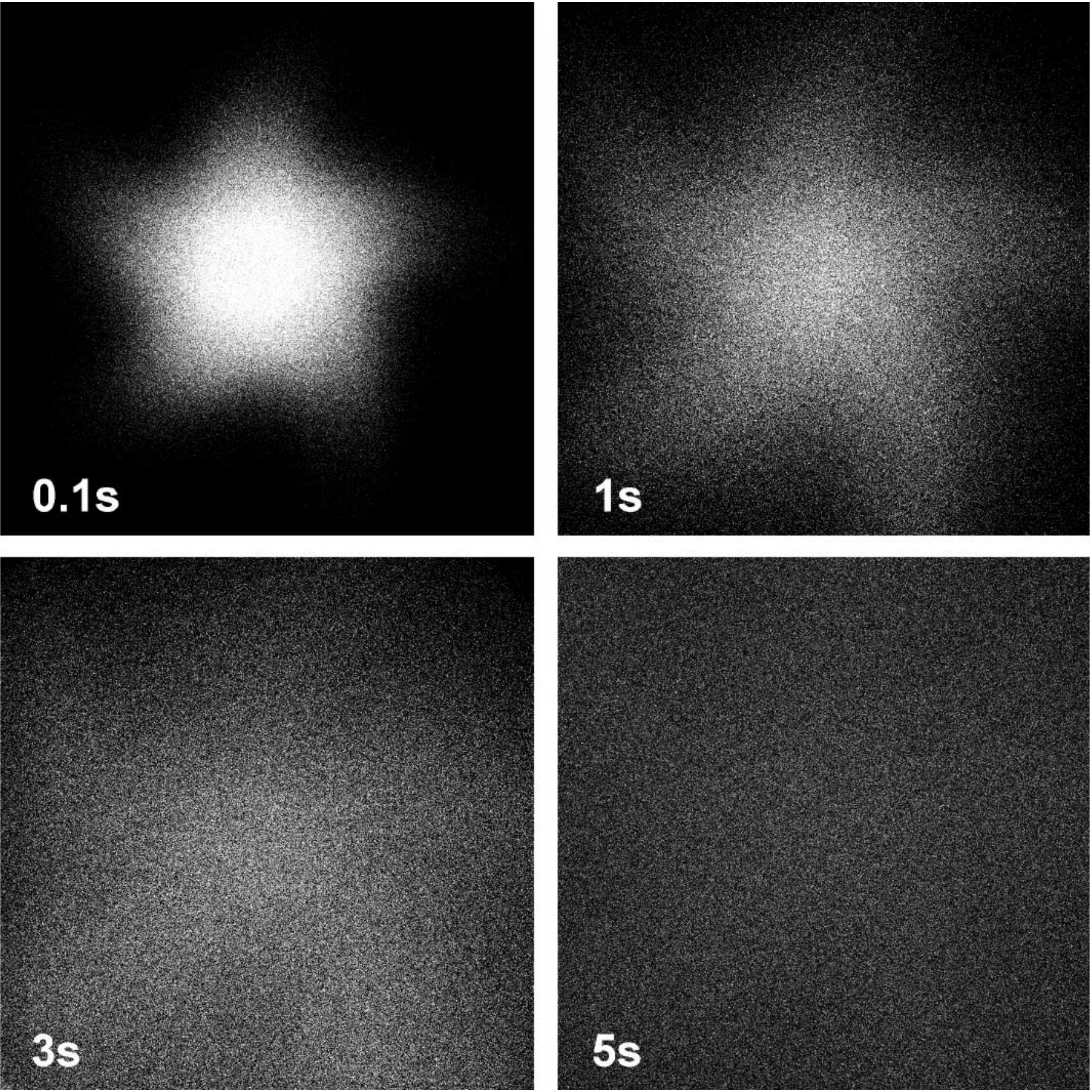}} \caption{The shutter illumination effect, as measured in domeflat images with short integration times of 0.1\,s (\emph{upper left}), 1\,s (\emph{upper right}), 3\,s (\emph{lower left}), and 5\,s (\emph{lower right}). Each image shows the full STK field of view. While the illumination pattern, induced by the limited speed of the STK shutter ($\sim$33\,ms), becomes significant for integration times shorter than 3\,s, it is not detectable anymore for integration times longer than 5\,s.} \label{shutter}
\end{figure}

Because of the limited speed of the STK shutter (the STK shutter fully opens within $\sim$33\,ms) a typical illumination pattern appears in shortly integrated images. This shutter illumination effect is show in Fig.\,\ref{shutter}, as measured in domeflat images with different short integration times.

\begin{figure*}[t!]
\centering
\resizebox{0.9\hsize}{!}{\includegraphics[]{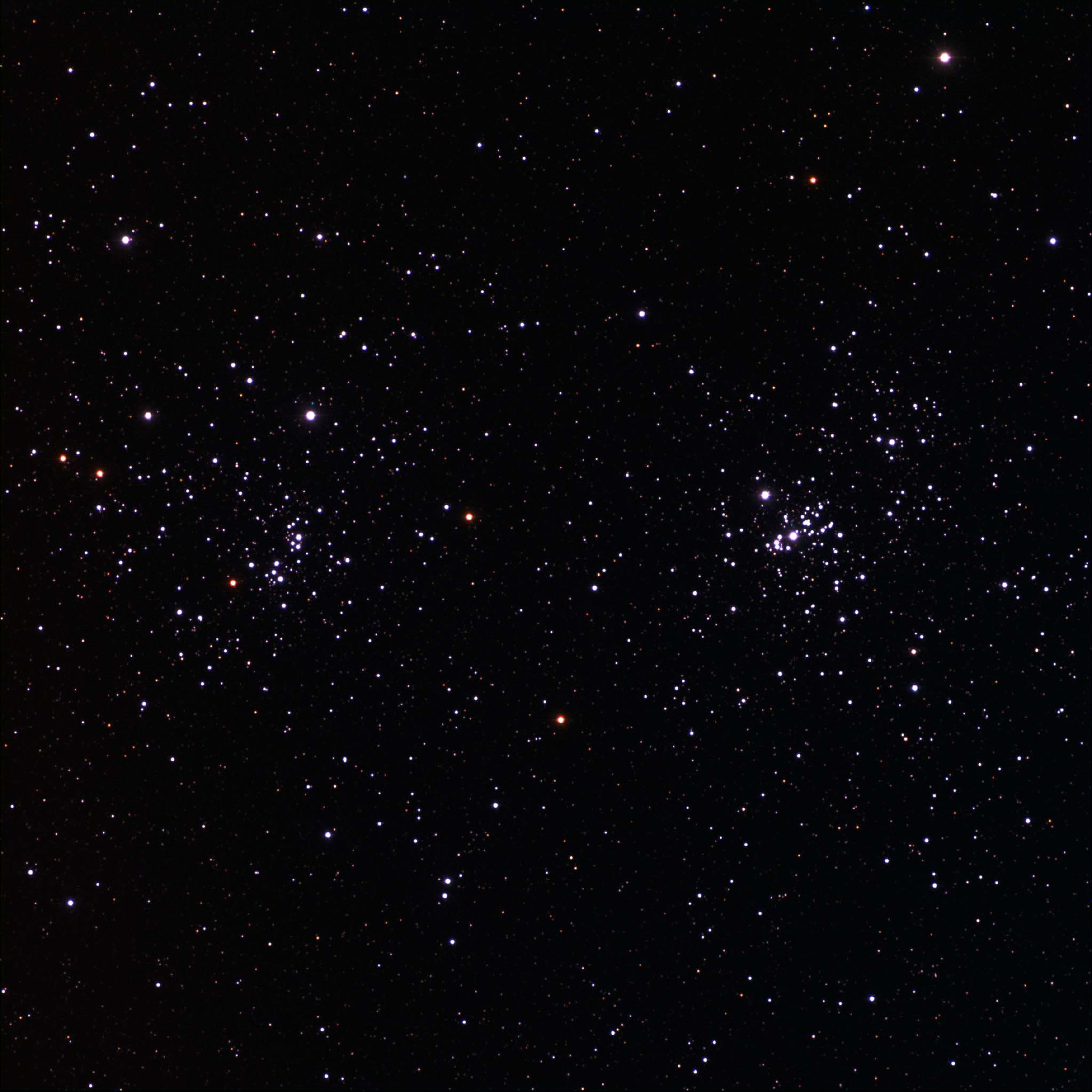}} \caption{The famous open clusters NGC869 and NGC884 (h\,+\,$\chi$\,Per) in the constellation Perseus. This image is the STK firstlight, taken on 2009 February 5. It is a color composite image, composed of three images taken through the STK $B$-, $V$-, and $R$-band filter, with integration times of 20\,s in the $B$-, and 10\,s each in the $V$- and $R$-band. The full STK field of view ($52.8'\times52.8'$) is shown with the North direction up and East to the left.}\label{firstlight}
\end{figure*}

For integration times shorter than 0.1\,s the intensity deviation between the border and the center of STK images is higher than 40\,\%. The five blade structure of the STK shutter is clearly visible in these shortly integrated images. The intensity deviation between the image border and its center decreases for longer integration times, namely to 3\,\% for integration times longer than 1\,s and 1\,\% for integration times longer than 3\,s. The shutter illumination effect is not detectable anymore in images with integration times  longer than 5\,s. Therefore, STK images with integration  times longer than 5\,s can be flatfielded with standard flatfield images, whereas for shorter integrated images special flats have to be taken with the corresponding short integration times.

\section{STK first light}

After its successful implementation in the Schmidt-focus of the 0.9\,m telescope at University Observatory Jena, the STK saw its first light in the night of 2009 Feb 5. The first light observations of the new camera were carried out under good atmospheric conditions. Three images of the famous open clusters h\,+\,$\chi$\,Per (NGC869 and NGC884) were taken through the STK $B$-, $V$-, and $R$-band filters, each after an optimized focusing of the camera. Integration times of 20\,s in the $B$-, as well as 10\,s each in the $V$- and $R$-band were chosen. At first, all images were dark and flatfield corrected. All reduced images were then aligned with cross-correlation to determine the relative offsets of the individual images. Finally, all images were combined to the $BV\!R$-composite image, which is shown in Fig.\,\ref{firstlight}.

\acknowledgements{We want to thank the staff of the mechanic and electronic division of the faculty for physics and astronomy at the University Jena, as well as the staff of 4pi Systeme GmbH for all their help and assistance during the construction and testing phase of the STK. Furthermore, we would like to thank the Thuringian State (Th\"{u}ringer  Ministerium f\"{u}r Bildung,  Wissenschaft und Kultur) in project number B 515-07010 for financial support. This publication makes use of data products from the Two Micron All Sky Survey, which is a joint project of the University of Massachusetts and the Infrared Processing and Analysis Center/California Institute of Technology, funded by the National Aeronautics and Space Administration and the National Science Foundation, as well as the \mbox{SIMBAD} and VIZIER databases, operated at CDS, Strasbourg, France.}

\end{document}